# 3×3 Slot waveguide crossing based on Maxwell's fisheye lens

S. Hadi Badri[a,*], M. M. Gilarlue[a], H. Soofi[b], H. Rasooli Saghai[c]

[a] Department of Electrical Engineering, Sarab Branch, Islamic Azad University, Sarab, Iran

[b] School of Engineering- Emerging Technologies, University of Tabriz, Tabriz 5166616471, Iran

[c] Department of Electrical Engineering, Tabriz Branch, Islamic Azad University, Tabriz, Iran

* sh.badri@iaut.ac.ir

## Abstract

Intersection of two or more silicon slot waveguides is inevitable in modern optical integrated circuits based on silicon on insulator (SOI) platform. In this article, we design a Maxwell's fisheye lens as the crossing medium for three Si slot waveguides and numerically investigate its characteristics such as insertion loss, crosstalk and bandwidth. For the 3×3 slot waveguide crossing, the average insertion loss of 1.2 dB and crosstalk levels lower than –15.1 dB are achieved in an ultra-wideband wavelength range of 415 nm covering the entire O, E, S, C, L, and U bands of optical communications. The footprint of the 3×3 silicon slot waveguide crossing presented in this article is merely 2×2 $\mu m^2$ which is considerably smaller compared to the previously designed Si slot waveguide crossings even with fewer number of ports. The proposed design can be expanded to support the intersection of more slot waveguides.

## Keywords

## 1. Introduction

As the number of components on the limited space of a modern optical integrated circuit increases, crossing of waveguides becomes inevitable which makes compact waveguide crossings indispensable components for highly dense optical and photonic integrated circuits. Therefore, various methods have been proposed in the literature to efficiently cross Si rectangular [1-5] and photonic crystal [6-10] waveguides. Si slot waveguides are key building blocks for silicon on insulator (SOI) platform due to their superior characteristics such as high light confinement and low loss. Due to these features, many devices are presented based on Si slot waveguides such as modulators [11, 12], sensors [13, 14], light sources [15, 16], demultiplexers [17], polarization splitters [18], all-optical logic gates [19], quantum optical circuits [20], and ring resonators [21]. However, only a few designs are presented so far for the crossing of Si slot waveguides. In [22], slot to strip mode converters followed by a multimode interference device is utilized to cross two slot waveguides resulting in an average insertion loss of approximately 0.1 dB and crosstalk levels of lower than -27 dB for a bandwidth of 200 nm. However, the footprint of the proposed device is considerably large equal to 15.6×15.6 µm². The same method with logarithmical mode converters are also studied [23]. In a bandwidth of 200 nm, the average insertion loss is about 0.2 dB whereas the crosstalk is lower than -33 dB. Nevertheless, the footprint of 20.8×20.8 µm² is the drawback of the proposed device again which makes it incompatible for highly dense integrated circuits. The transmission efficiency can also be improved by filling up the crossing slots locally with an insertion loss of approximately 1.1 dB [24]. Vertical coupling is another method presented to reduce the radiation loss in the crossing of Si slot waveguides, however, the large footprint of 13.8×0.43 µm² [25] makes it unsuitable for compact photonic integrated circuits. It is worth noting that all the above methods only support the crossing of two waveguides with the crossing angle of 90°.

In this paper, we design an intersection of three Si slot waveguides based on the imaging properties of the Maxwell's fisheye (MFE) lens. Radiation of the point source on the surface of the lens is focused on the diametrically opposite point of the MFE lens. To the best of our knowledge, it is the first time that a 3×3 Si slot waveguide crossing is designed and numerically investigated. Moreover, the proposed method can be easily expanded to increase the number of intersecting waveguides by increasing the radius of the lens. The refractive index of the MFE lens is

$$n_{lens}(r) = \frac{2 \times n_{edge}}{1+(r/R_{lens})^2} \quad , \quad (0 \leq r \leq R_{lens}) \tag{1}$$

In Eq. (1), $R_{lens}$ is the radius of the lens and $r$ is the radial distance from the center of the lens, and $n_{edge}$ is the refractive index of the lens at its edge. Recently, interesting applications have been introduced for gradient index lenses such as MFE [26, 27], Luneburg [28, 29], and Eaton [30] lenses.

## 2. Simple intersection of three slot waveguides

In this article, a Si slot waveguide with slot width and height of $w_s$=100 nm and $h$=250 nm is considered. The width of the Si rails is $w_h$=200 nm. Fig. 1 illustrates the contour plot of the quasi-TE mode at λ=1550 nm for the slot waveguide where silicon rails are surrounded by silica. $E_x$ is the main electric field component which is symmetric about the $y$ axis and has a large discontinuity resulting in highly enhanced field in the slot. The guided mode displayed in Fig. 1 is confined in the slot by total internal reflection and consequently there are no confinement losses [22]. The electric field distribution for a simple intersection of three slot waveguides at 1550 nm is shown in Fig. 2. Transmission, crosstalk, and reflection values are also displayed in this figure. The diffraction at the simple intersection is considerably large leading to substantial insertion loss of 11.2 dB while the crosstalk in the other ports is as high as -4.7 dB.

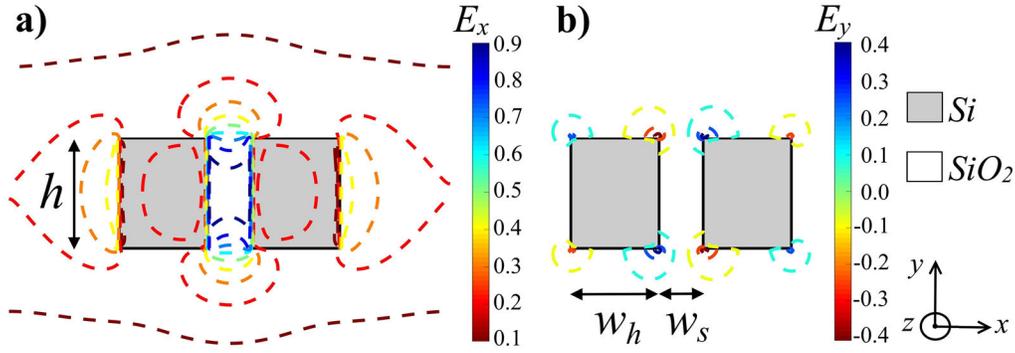

Fig. 1. The contour plot of the quasi-TE mode at λ=1550 nm for the slot waveguide. The silicon rails are surrounded by silica.

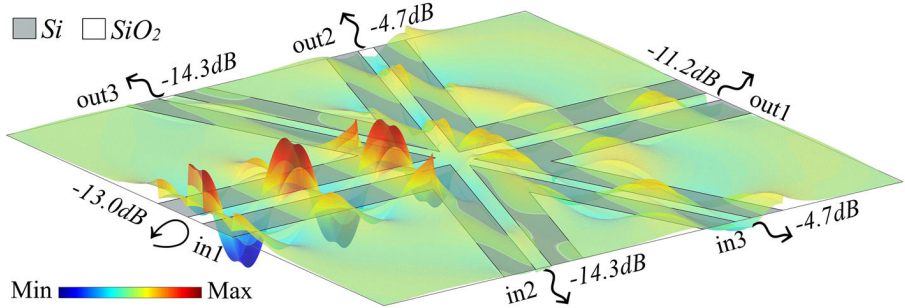

Fig. 2. The electric field distribution for a simple intersection of three slot waveguides at 1550 nm.

## 3. Multilayered MFE lens design

When the width of the layers in the multilayer structure is comparable to the wavelength of the incident light, the interference is the dominant phenomenon in determining the response of the structure. Various devices have been designed based on the interference effect in multilayer structures [31-33]. However, when the width of each layer is much smaller than the wavelength, the interference effect becomes negligible and the multilayer structure can be treated as an anisotropic metamaterial medium governed by effective medium theory [6, 34-39]. We implement the MFE lens based on this concept. Due to the symmetry of the refractive index profile of the MFE lens, it can be realized by a concentric-ring multilayer structure [6, 39]. When the width of each layer is much smaller than the wavelength, the multilayer structure can be regarded as an effective medium [38]. Here, silicon and silica are considered as the constituting materials of the concentric-ring multilayer structure at which Si (SiO$_2$) serves as the host (inclusion) material. The two components of the permittivity tensor, $\varepsilon_\parallel$ and $\varepsilon_\perp$, depend on the parallel or perpendicular arrangement of the inclusion layers with respect to the direction of the electric field, and consequently have different values. In practice, the electric field is not purely TE or TM mode, therefore, the effective permittivity depends on both of these components. We simplify the design procedure by only considering $\varepsilon_\perp$ component [40] where the inclusion layers are perpendicular to the electric field [41]

$$\varepsilon_{eff}^{TE} = \frac{\varepsilon_{host}\varepsilon_{inc}}{\varepsilon_{inc}f_{inc} + \varepsilon_{host}(1-f_{inc})} \qquad (2)$$

where $\varepsilon_{eff}^{TE}$ is the effective permittivity of the cell for TE mode and $\varepsilon_{host}$ and $\varepsilon_{inc}$ are the permittivities of the host and inclusion materials respectively. The filling factor, $f_{inc}$, is the fraction of the total volume occupied by the inclusion layer. In our design, the lens is divided into equal segments with a width of $\Lambda$. Afterwards, the average refractive index in each layer ($\varepsilon_{eff}^{TE}$) is calculated. Subsequently, the width of the inclusion layer in the i-th layer ($d_i$) is calculated. To do so, Eq. (2) is rearranged to give the filling factor for the effective permittivity as

$$f_{inc} = \frac{\varepsilon_{host}(\varepsilon_{inc} - \varepsilon_{eff}^{TE})}{\varepsilon_{eff}^{TE}(\varepsilon_{inc} - \varepsilon_{host})} \qquad (3)$$

The filling factor for the i-th layer is $f_{inc,i} = A_{inc,i}/A_i$ where $A_{inc,i} = 2\pi r_i d_i$ and $A_i = 2\pi r_i \Lambda$ are the areas of the inclusion (in the i-th layer) and the i-th layer itself, respectively. $r_i$ is the distance from the origin to the middle of the i-th layer. Therefore, $d_i$ can be obtained by

$$d_i = \frac{\varepsilon_{host}(\varepsilon_{inc} - \varepsilon_{eff}^{TE})}{\varepsilon_{eff}^{TE}(\varepsilon_{inc} - \varepsilon_{host})} \Lambda \qquad (4)$$

The designing concept is shown in Fig. 3(a). Only a quarter of the annular rings is shown in this figure. The designed lens is illustrated in Fig. 3(b). However, the width of outer layers is less than 4 nm which may not be feasible, hence, the width of the inclusion layers is limited to a more feasible value. As shown in Fig. 3(c), the inner and outer layers are limited to 35 nm and 18 nm respectively. This was achieved by dividing each annular ring into annular sectors. The spacing between the center of two consecutive annular sectors was considered to be the same as the period of the structure ($\Lambda = 77.5\ nm$). Then in each annular sector we placed an inclusion with the shape of annular sector with the desired width. Similar to graded photonic crystals (GPC), the inclusions can also be in the shape of cylindrical rods. Afterwards, the arc length of the inclusion sector is calculated to satisfy Eq. (3). Although in this article, the lens is implemented by the multilayer structure, however, other methods such as varying the thickness of guiding layer and GPC can also be utilized to realize GRIN lenses [42].

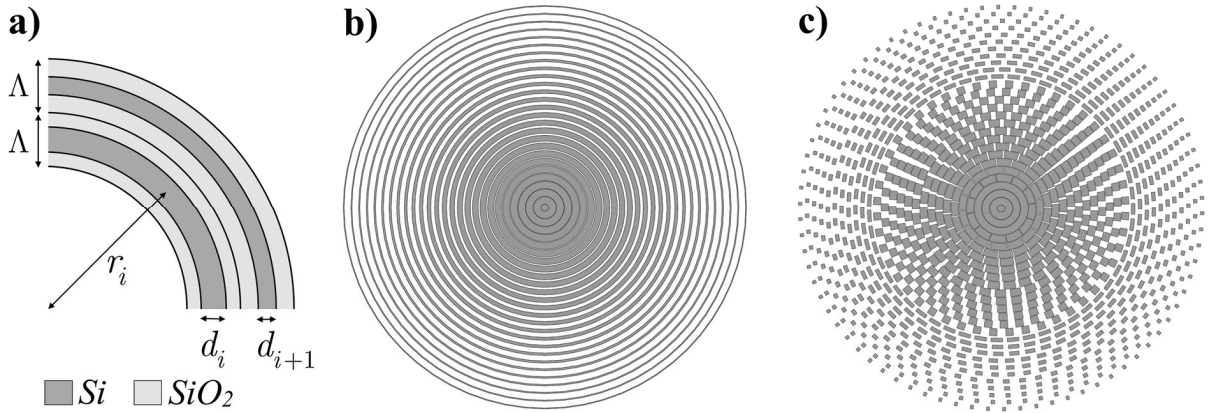

Fig. 3. a) The designing concept for cylindrical multilayer structure. b) Simple cylindrical multilayer structure where the outer layers are very thin. c) The width of inclusion layers is limited a minimum value so the arc length of inclusion layers is controlled to satisfy Eq. (3). The host material is not shown in b) and c).

## 4. Results and discussion

In this article, a two-dimensional finite-difference time-domain (FDTD) is utilized to evaluate the performance of the proposed Si slot waveguide crossing. Such a modeling procedure is proven to be accurate for similar structures previously [22, 23, 43]. The built-in material models of silicon and silica of the Lumerical software are used in the simulations. A mode source is used to inject a TE mode into the simulation region where the maximum meshing step is 5 nm. The electric field distribution of the 3×3 slot waveguide crossing is shown in Fig. 4 at a wavelength of $\lambda$=1550 nm. At this wavelength, the insertion loss is reduced from 11.2 dB (for the case of simple 3×3 crossing) to 1.1 dB by the help of the MFE lens. Crosstalk to other ports is also reduced from -4.7 dB to -16.0 dB. The designed crossing has small footprint of 2×2 $\mu m^2$. In order to compare the imaging quality of the designed MFE lens, the electric field intensity profiles at the input and output of the lens are compared in Fig. 5.

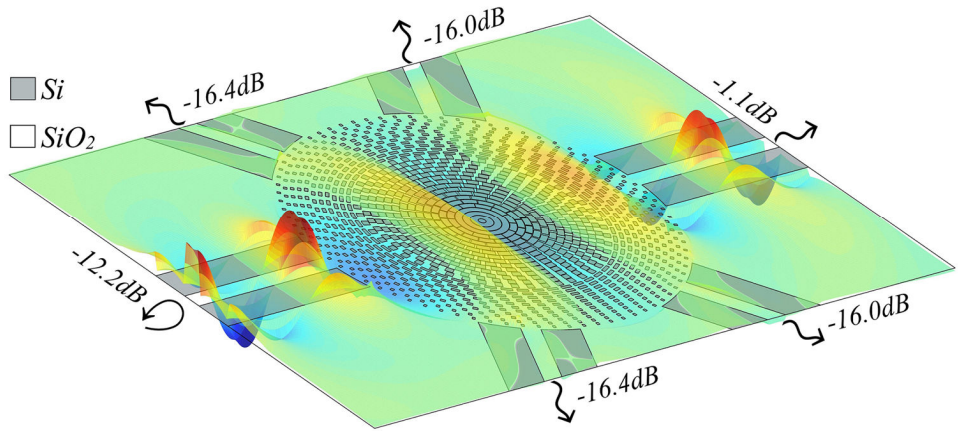

Fig. 4. The electric field distribution of the 3×3 slot waveguide crossing based on the MFE lens at a wavelength of 1550 nm.

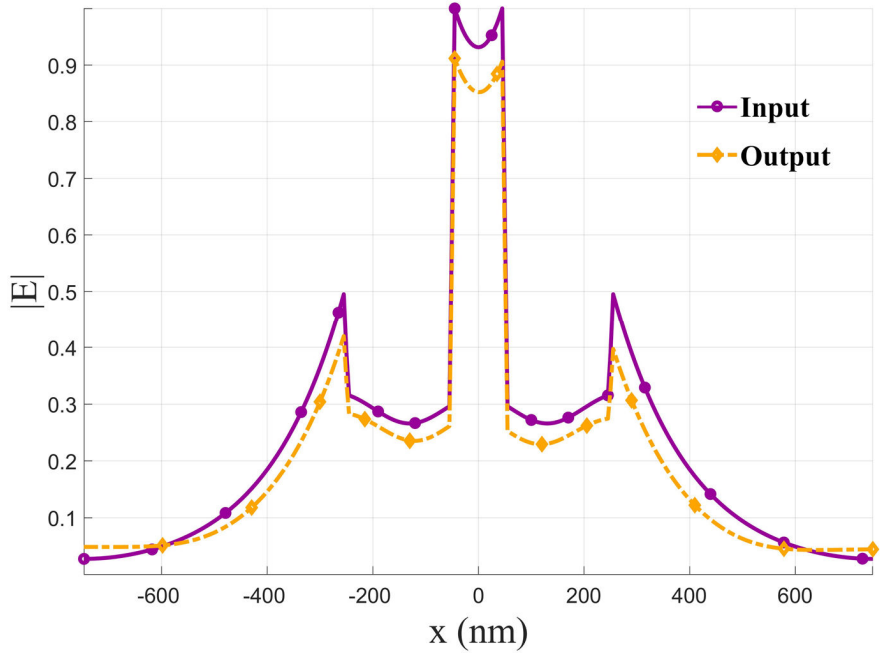

Fig. 5. Electric field intensity at the input and output of the designed 3×3 slot waveguide crossing.

The scattering parameters of the 3×3 slot waveguide crossing is illustrated at Fig. 6. Since the electric field is confined in a subwavelength region in the slot waveguide, satisfying the effective medium condition requires the period of the structure to be smaller than the width of the slot. Hence, the period of the structure is chosen to be $\Lambda = 77.5\ nm$. Simulations reveal that for $\Lambda > w_s$ the reflection in the interface of the slot waveguide and the MFE lens structure is considerably high. In the O-band, the return loss is as high as -7 dB and consequently the insertion loss increases to 1.9 dB. The lowest insertion loss, 0.76 dB is obtained at λ=1435 nm at which the return loss reaches its minimum value. In the C-band, the average insertion loss is 1.16 dB and the crosstalk levels are lower than -16.0 dB. The return loss in this band is lower than -12 dB. For the entire O, E, S, C, L, and U bands of optical communications, the crosstalk levels are lower than -15.1 dB.

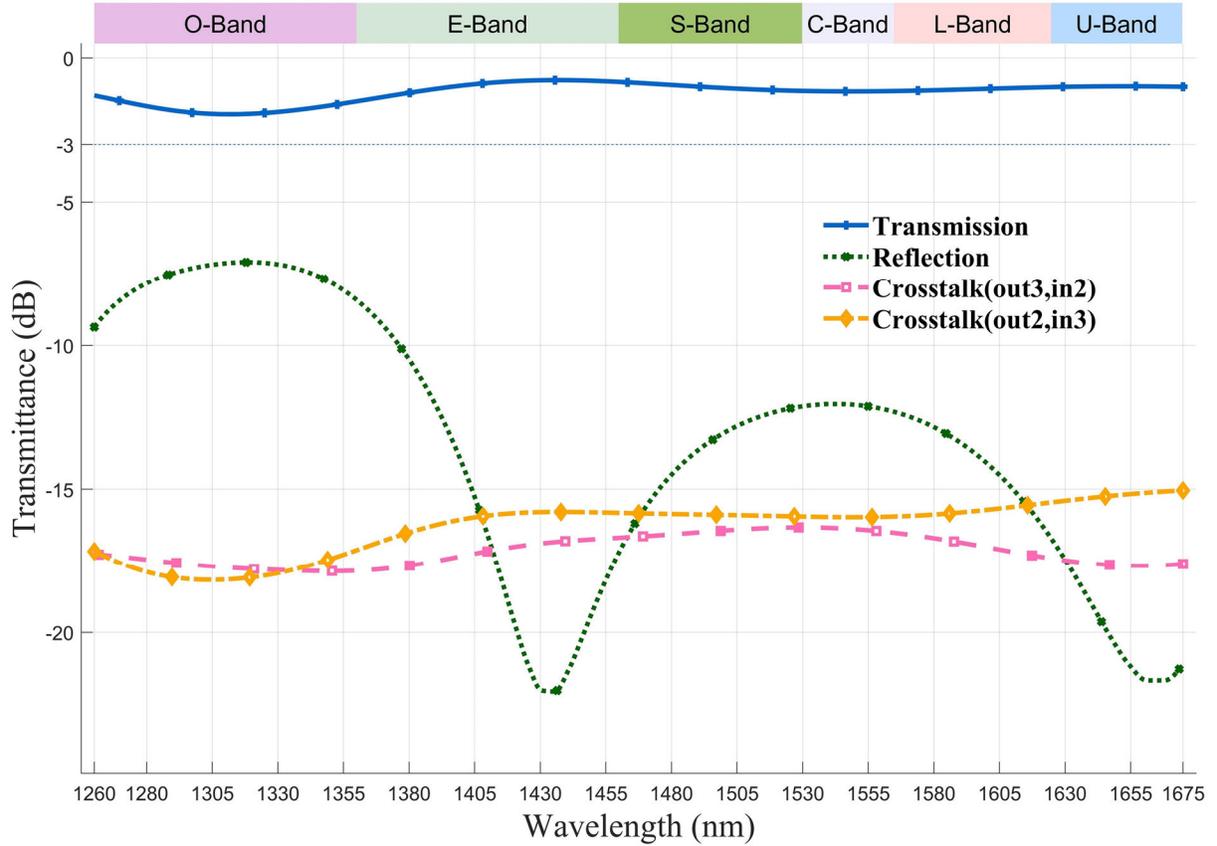

Fig. 6. The scattering parameters of the 3×3 slot waveguide crossing based on the MFE lens.

Finally, it is a good practice to compare the performance characteristics of the designed lens with other Si slot waveguide crossings. The insertion loss and crosstalk of previous studies [22, 23, 25] are better relative to the design presented in this article. However, these designs only offer a solution for 2×2 slot waveguide crossings whereas in this article a 3×3 waveguide crossing is proposed. Moreover, the presented device can be simply expanded to support higher number of waveguide crossings merely by increasing the size of the lens. Moreover, the footprint of our design is 2×2 μm$^2$ while the designs of [22, 23, 25] have a considerably larger footprint. Other methods only offer a crossing angle of 90° with small deviations whereas the crossing angle is a flexible parameter in our design. Last but not least, the design presented in this article has the broadest bandwidth reported so far for Si slot waveguide crossings.

## 5 Conclusion

In this article, a 3×3 Si slot waveguide crossing is proposed for the first time and numerically investigated by FDTD. We have successfully decreased the insertion loss of the simple intersection of three slot waveguides from 11.2 to 1.2 dB by utilizing the MFE lens as crossing medium. The proposed crossing is implemented by ring-based multilayer structure with a small footprint of 2×2 μm$^2$ which is considerably small compared to the available 2×2 crossings. The average insertion loss is 1.2 dB and crosstalk levels are below -15.1 dB in the entire O, E, S, C, L, and U bands of optical communication.